\documentclass{article}
\usepackage[utf8]{inputenc}
\usepackage{graphicx}
\usepackage{amsmath}
\usepackage{amsmath,bm}
\usepackage{cite}
\usepackage{appendix}
\usepackage{authblk}

\title{Optical needles with arbitrary homogeneous  three-dimensional   polarization}

\author{Li Hang}
\author{Ying Wang}
\author{Peifeng Chen \thanks{Corresponding author: pfchen@hust.edu.cn \ Li Hang: lihang5011@vip.qq.com}}

\affil{School of Optical and Electronic Information, Huazhong University of Science and Technology, Wuhan, China}

\date{}

\begin{document}
\bibliographystyle{unsrt}
\maketitle

\begin{abstract}
    we propose a new method to generate optical needles by focusing vector beams comprised of radially polarized component and azimuthally polarized vortex components. The radial part can generate longitudinal polarization, while the azimuthal parts can generate left- and right-handed polarization. Hence, an arbitrary 3D polarization can be obtained. To our knoeledge, it may be the first time that    arbitrarily polarized optical  needles whose transverse sizes are under 0.5$\lambda$ have been achieved. And their polarized homogeneity is beyond 0.97.
\end{abstract}

\section{Introduction}
Over the past few years, there are several papers about generating optical fields with three-dimensional   polarization by tightly focusing \cite{abouraddy2006three, chen2010diffraction,zhu2013generation}.  However, they do not show a quantitative relationship between the input field and the three-dimensional   polarization of the focal field. 
In this paper, we propose a new method to generate super-resolved optical fields by focusing complex input field. Such optical fields have large longitudinal full  width at half maximum (LFWHM) and small transverse full  width at half maximum (TFWHM), so they are called optical needles \cite{panneton2015needles,hang2018needles}. The input field comprise radially polarized beam and azimuthally polarized vortex beams. The radially polarized beam  can generate the longitudinally polarized field \cite{zhan2009cylindrical}, while the azimuthally polarized  beams with vortex phase $\exp(\pm i\varphi)$ can generate left- and  right-handed polarized fields. These three components constitute three-dimensional unitary bases. Thus, an arbitrary polarization can be obtained.

\section{Theory and configuration}

\begin{figure}[htbp]
	\centering
	\includegraphics{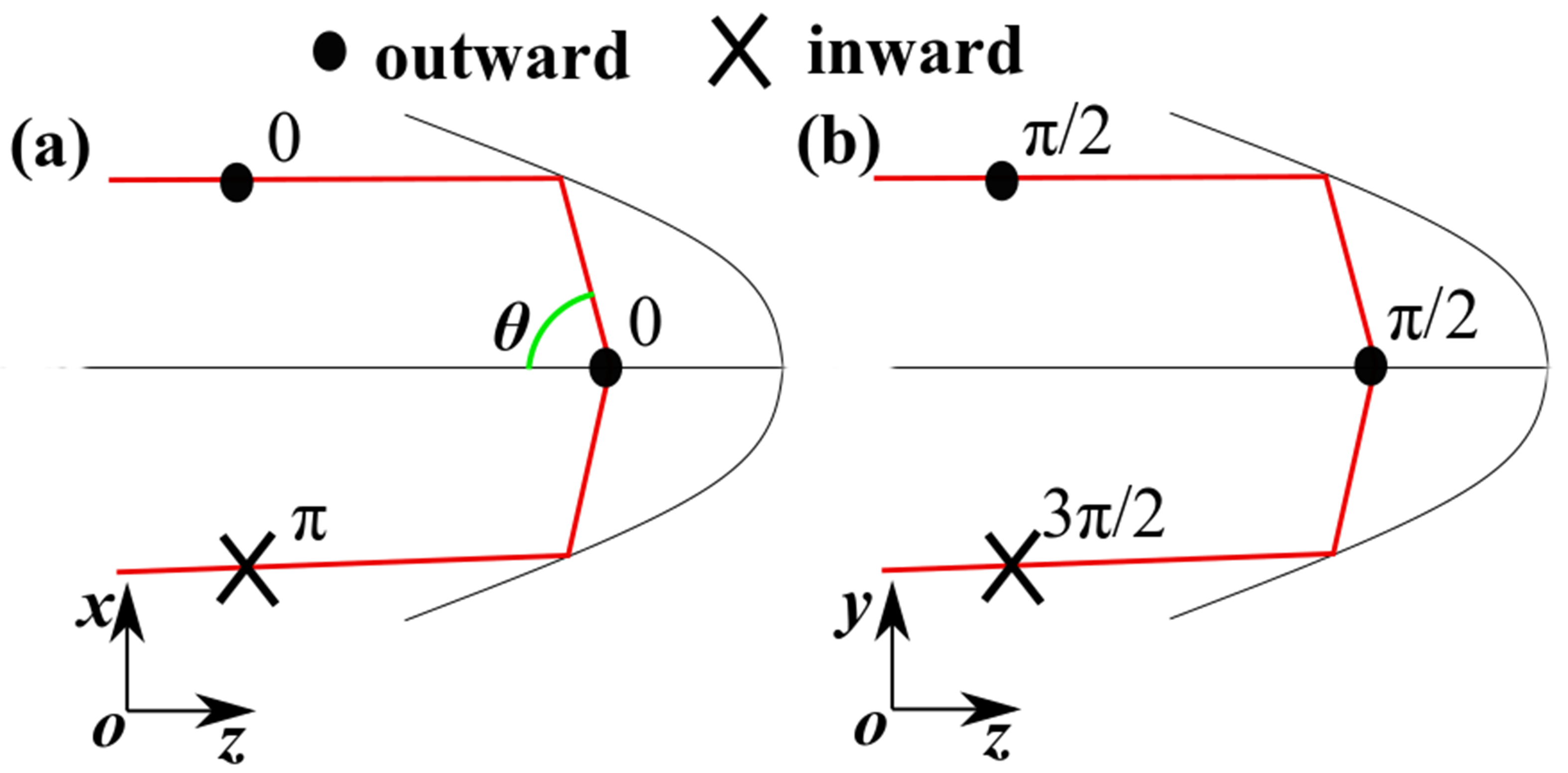}
	\caption{Schematic diagrams of an azimuthally polarized beam with vortex phase $\exp(i\varphi)$  generating a left-handed polarized beam. (a) in $xOz$ plane;  (b) in $yOz$ plane.}
	\label{Fig:setup}

\end{figure}

Figure \ref{Fig:setup} shows how an azimuthally polarized  beam with vortex phase $\exp(i\varphi)$ generates a left-handed polarized beam. The dots and crosses are the directions of the electric field when the phases are not considered. The numbers near the dots and crosses are the phases of the electric fields. In Fig. \ref{Fig:setup}(a), the electric field near the focus is parallel to $y$-axis, while in Fig. \ref{Fig:setup}(b) it is parallel to $x$-axis. Since there is a phase retardation of $\pi/2$, the total field is left-handed polarized. Similarly, the focal field will be right-handed polarized when the incident beam is azimuthally polarized  with vortex phase $\exp(-i\varphi)$. In fact, this is an example of orbital-to-spin conversion \cite{yu2018orbit,Hang:19}. An annular input beam comprised of radially polarized component and azimuthally polarized vortex can be expressed as 

\begin{equation}
\begin{aligned}
\mathbf{E}_{in}=\left\{
\begin{aligned}
 & =f\mathbf{e}_r+g_{-1}\exp(-i\varphi)\mathbf{e}_\varphi+g_{1}\exp(i\varphi)\mathbf{e}_\varphi ,& |\theta-\theta_0|\leq \Delta\theta/2\\
 & =0, & |\theta-\theta_0|> \Delta\theta/2, 
\end{aligned}
\right.
\end{aligned}
\end{equation}
where $\theta_0$ and $\Delta\theta$ are the angular position and width of the incident beam. Such beams can be generated by spatial light modulators \cite{han2013vectorial,rosales2017simultaneous}. For a parabolic mirror in Fig. \ref{Fig:setup},  $\theta_0=90^\circ$ is chosen in this paper. $f$, $g_{-1}$ and $g_1$ are complex numbers. According to the  Richards-Wolf vector diffraction theory \cite{richards1959electromagnetic}, when $\Delta\theta\ll \theta_0$, the electric field near the focus can be expressed as \cite{hang2019theoretical}

\begin{equation}
\begin{aligned}
\mathbf{E}(r,\phi,z)\approx \frac{A(z;\theta_0,\lambda)}{\sqrt2}   \left[\begin{array}{ccc}-i g_{-1}(e^{-2i\phi}J_{2}+J_{0})\mathbf{e}_x+ig_{1}(e^{2i\phi}J_{2}+J_{0})\mathbf{e}_x \\  g_{-1}(e^{-2i\phi}J_{2}-J_{0})\mathbf{e}_{y}+ g_1(e^{2i\phi}J_{2}-J_{0})\mathbf{e}_{y} \\2 fJ_{0}\mathbf{e}_z \end{array}\right]\\ = A(z)   \left[\begin{array}{ccc} i(-g_{-1}J_{0}+g_1e^{2i\phi}J_{2})\mathbf{e}_R \\  i(g_1J_{0}-g_{-1}e^{-2i\phi}J_{2})\mathbf{e}_{L} \\\sqrt{2} fJ_{0}\mathbf{e}_z \end{array}\right], 
\label{Eq:needle}
\end{aligned}
\end{equation}
where $A(z)\propto \frac{\sin(kz\sin \frac{\delta\theta}{2})}{kz} $, $k$ is wave number. $J_n=J_n(kr)$, where $J_n(\cdot)$ denotes the nth-order Bessel function of the first kind. $\mathbf{e}_R=(\mathbf{e}_x-i\mathbf{e}_y)/\sqrt{2}$ and $\mathbf{e}_L=(\mathbf{e}_x+i\mathbf{e}_y)/\sqrt{2}$ represent the unit vector of right- and
left-handed circular polarization states, respectively.  Obviously, the focal filed is a non-diffracting beam. Its LFWHM is $0.8713\lambda/\Delta\theta$ \cite{hang2018tunable,hang2019theoretical}. Its TFWHM can be obtain by solving the equation
\begin{equation}
   |\mathbf{E}(r,\phi,z)|^2=|\mathbf{E}(0,0,z)|^2/2.
   \label{Eq:TFWHM}
\end{equation}
The TFWHM is dependent on $\phi$ and the numerical results are shown in the next section.
The electrical field strength at the focus is $\mathbf{E}_0=\mathbf{E}(0,0,0)=-ig_{-1}\mathbf{e}_R+ig_1\mathbf{e}_L+\sqrt{2}f\mathbf{e}_z$ ($A(z)$ will be omitted in the following).  To evaluate the polarized homogeneity near the focus quantitatively, we can define polarized homogeneity as
\begin{equation}
    \eta=\frac{\int_S|\mathbf{E}^*\cdot \frac{\mathbf{E}_0}{|\mathbf{E}_0|}|^2dS}{ \int_S\mathbf{E}^*\cdot \mathbf{E}dS},
\end{equation}
where $S$ is the  integral area in the focal plane. Considering a circular area $r<R$, the polarized homogeneity can be simplified as

\begin{equation}
    \eta=\frac{\int_0^R [(\alpha^2+\beta^2+2\gamma^2)^2J_0^2+2\alpha^2\beta^2J_2^2] rdr}{\int_0^R[(\alpha^2+\beta^2+2\gamma^2)^2J_0^2+(\alpha^2+\beta^2+2\gamma^2)(\alpha^2+\beta^2)J_2^2] rdr}
    \label{Eq:PURE}
\end{equation}

with

\begin{equation}
    \begin{aligned}
    \alpha &=|g_{-1}|/\sqrt{|g_{-1}|^2+|g_1|^2+|f|^2}\\
    \beta &=|g_1|/\sqrt{|g_{-1}|^2+|g_1|^2+|f|^2}\\
    \gamma &=|f|/\sqrt{|g_{-1}|^2+|g_1|^2+|f|^2},
    \label{Eq:factor}
    \end{aligned}
\end{equation}
where $\alpha$, $\beta$ and $\gamma$ represent the  weighting factors of the three components of the incident beam. Only when $\alpha=\beta=0$, $\eta=1$. This is because   the azimuthally polarized beams with vortex phase $\exp(\pm i\varphi)$ cannot be transformed into left- or right-handed polarized beams completely. 
\section{Numerical simulation and discussion}

\subsection{TFWHM}

\begin{figure}[htbp]
	\centering
	\includegraphics{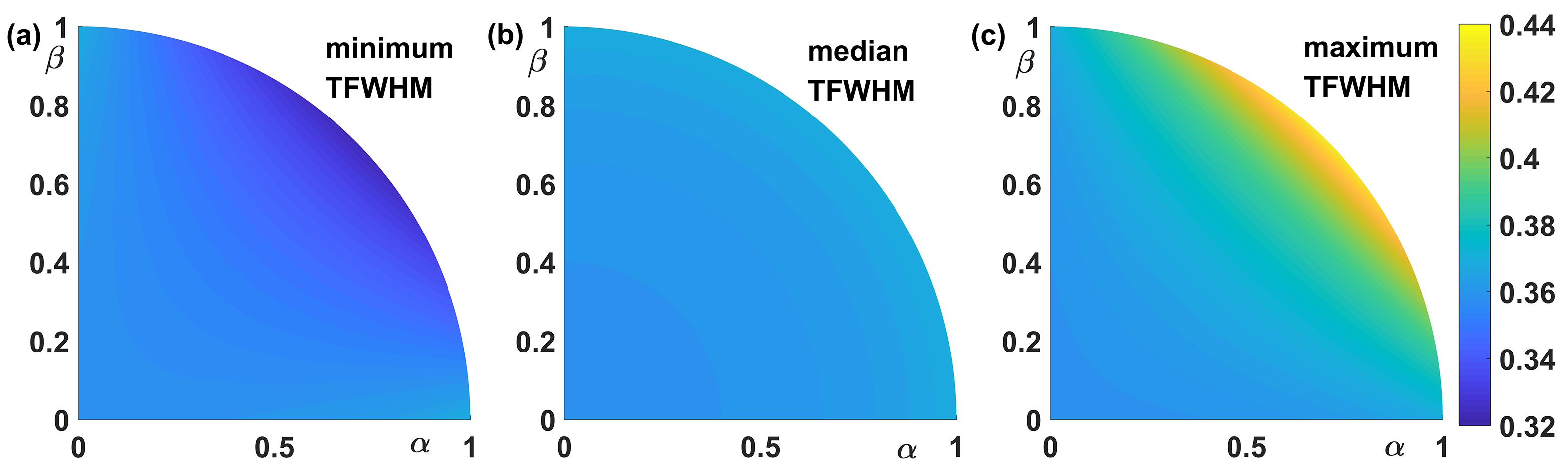}
	\caption{Minimum (a), median and maximum (c) TFWHM as  functions of $\alpha$ and $\beta$.}
	\label{Fig:TFWHM}
\end{figure}

Substituting  Eq. (\ref{Eq:needle}) and Eq. (\ref{Eq:factor}) into Eq. (\ref{Eq:TFWHM}),

\begin{equation}
\begin{aligned}
(\alpha^2+\beta^2+2\gamma^2)J_0^2+(\alpha^2+\beta^2)J_2^2-2\alpha\beta\cos(\delta_1-\delta_2-2\phi)J_0J_2 \\=(\alpha^2+\beta^2+2\gamma^2)/2,
\label{Eq:TFWHM2}
\end{aligned}
\end{equation}
where $\delta_1$ and $\delta_2$ are the complex arguments of $g_{-1}$ and $g_1$, respectively. Equation \ref{Eq:TFWHM2} shows that TFWHMs will be dependent on $\phi$ unless $\alpha\beta=0$. Because of the interaction of the two azimuthally polarized components, the focal spot is not a circle, but like an ellipse. Hence, we use minimum (when $\delta_1-\delta_2-2\phi=\pi$), median (when $\delta_1-\delta_2-2\phi=\pi/2$) and maximum (when $\delta_1-\delta_2-2\phi=0$) TFWHM to evaluate the size of focal spots. As shown in Fig. \ref{Fig:TFWHM}, their ranges are 0.32 $\sim$ 0.37$\lambda$, 0.36 $\sim$ 0.37$\lambda$ and 0.36 $\sim$ 0.43$\lambda$, respectively. When $\alpha^2=\beta^2=0.5$, the influence of the two azimuthally polarized components reaches the maximum, so the focal spot has the minimum value of minimum TFWHMs ($0.32\lambda$) and the maximum value of maximum TFWHMs ($0.43\lambda$). Median TFWHMs are only dependent on $\gamma$. The radially polarized component only generate zero-order Bessel component, while  the azimuthally polarized components can both  generate zero- and second-order Bessel component. Therefore, the median TFWHM increase as $\gamma$ decrease.

\subsection{Polarized homogeneity}
\begin{figure}[htbp]
	\centering
	\includegraphics{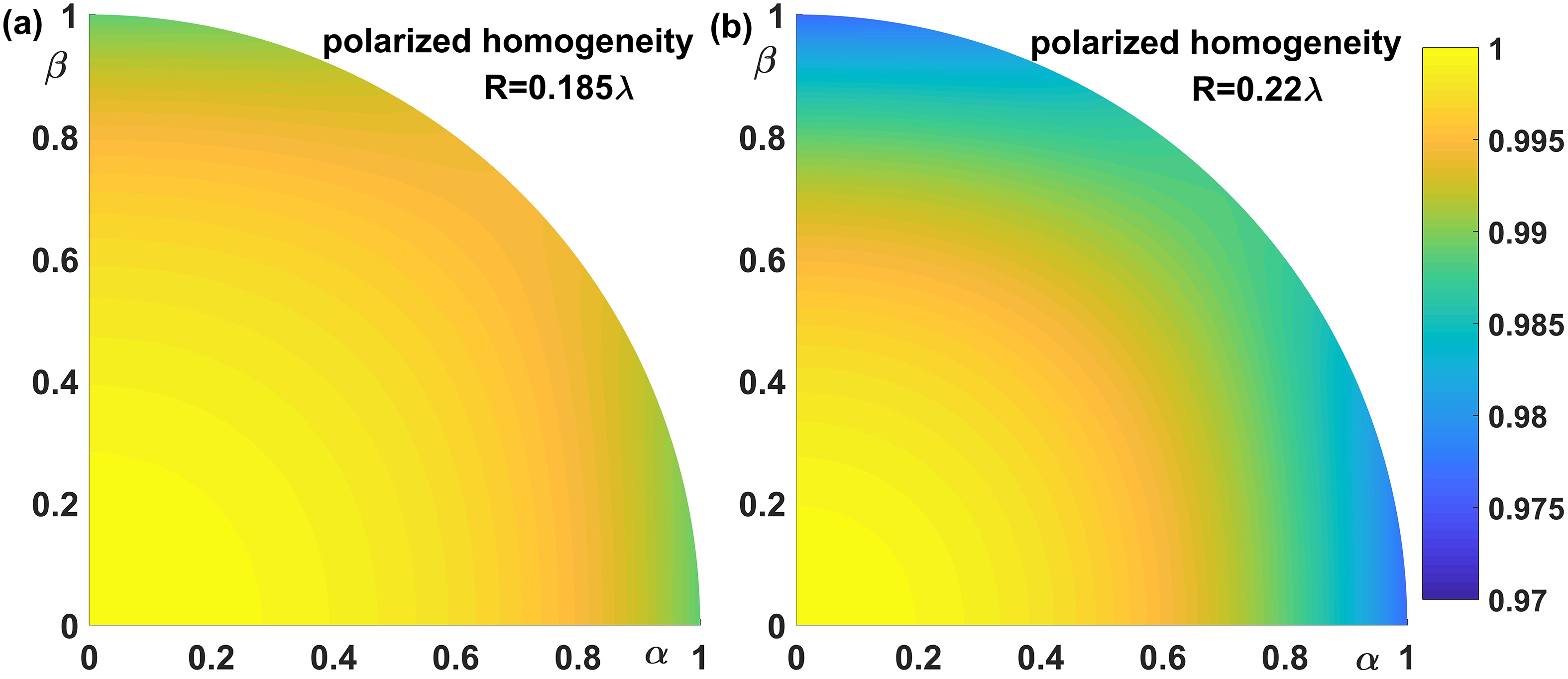}
	\caption{Polarized  homogeneity  as  functions of $\alpha$ and $\beta$. (a) and (b) correspond to median and maximum TFWHM, respectively.}
	\label{Fig:PURE}
\end{figure}

According to Eq. (\ref{Eq:PURE}), the polarized homogeneity is shown in Fig. \ref{Fig:PURE}. It is always larger than 0.989 when $R=0.185\lambda$ and larger than 0.977 when $R=0.22\lambda$. It increases as $\gamma$ increases because of the incomplete transformation of the azimuthally polarized components. For a fixed $\gamma$, the maximum polarized homogeneity is at $\alpha=\beta$. This is because the bad transformation of two azimuthally polarized components can  cancel partly.

\subsection{Examples}

\begin{figure}[htbp]
	\centering
	\includegraphics{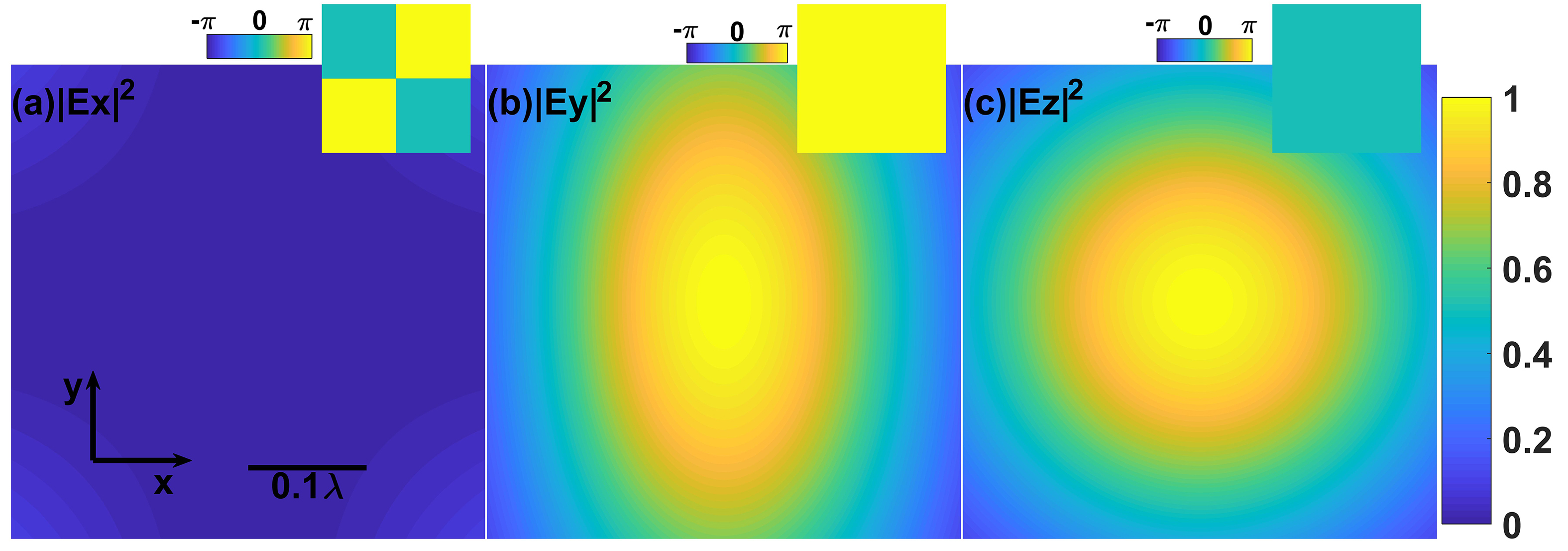}
	\caption{Intensity and phase (insets) profiles on the
focal plane when $g_{-1}=g_1=f$.}
	\label{Fig:LP}
\end{figure}

That $\mathbf{E}_0$ is linearly polarized is equivalent to $\mathbf{E}_0\times\mathbf{E}_0^*=\mathbf{0}$. Figure \ref{Fig:LP} is an example when $g_{-1}=g_1=f$. The $x$-component of electric field is nearly zero and the $y$-component and $z$-component have a  phase retardation of $\pi$. Hence, the focal field is linearly polarized in $yOz$ plane. The minimum, median and maximum TFWHM are 0.34, 0.36 and 0.39$\lambda$, respectively. The polarized homogeneity is 0.991 (R=0.22$\lambda$).

\begin{figure}[htbp]
	\centering
	\includegraphics{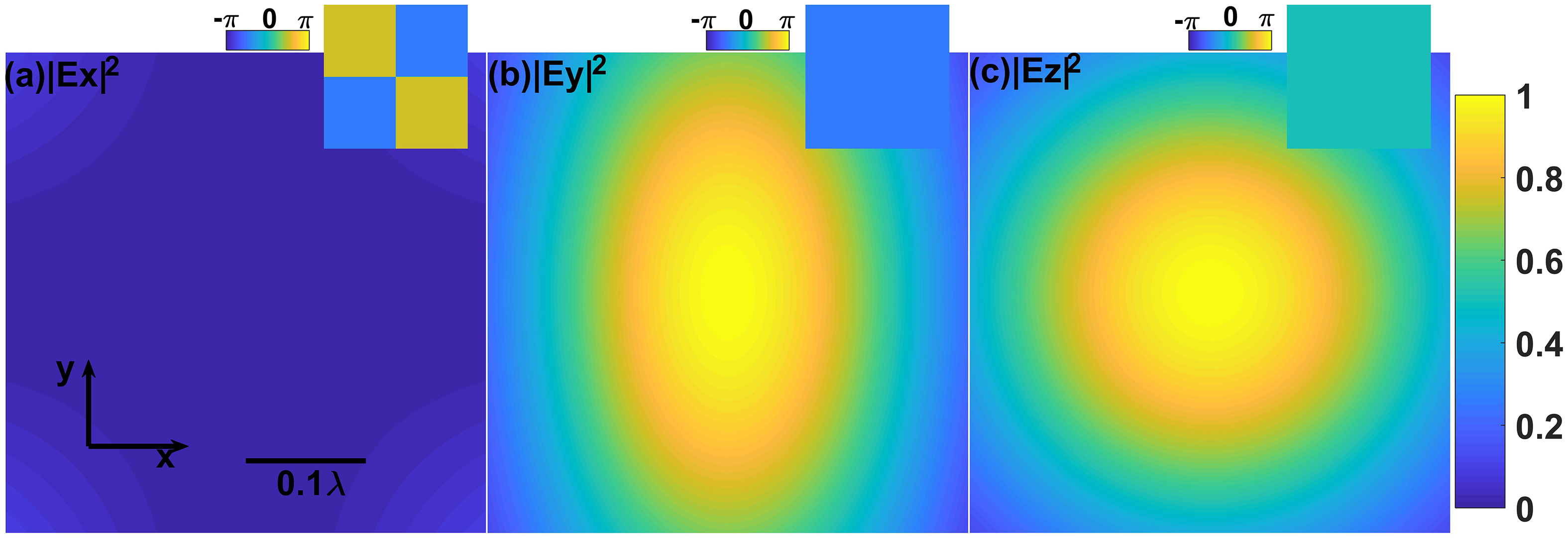}
	\caption{Intensity and phase (insets) profiles on the
focal plane when $g_{-1}=g_1=if$.}
	\label{Fig:CP}
\end{figure}

That $\mathbf{E}_0$ is circularly polarized is equivalent to $|\mathbf{E}_0\times\mathbf{E}_0^*|=\mathbf{E}_0\cdot\mathbf{E}_0^*$.  Figure \ref{Fig:CP} is an example when $g_{-1}=g_1=if$. Compared with last example, the azimuthally polarized components have phases of $\pi/2$. The intensity profiles in Fig. \ref{Fig:CP} are the same as Fig. \ref{Fig:LP}, while the phase retardation between  y-component  and  z-component  of the focal filed is $\pi/2$. Hence, the focal field is circularly polarized in $yOz$ plane. 

\begin{figure}[htbp]
	\centering
	\includegraphics{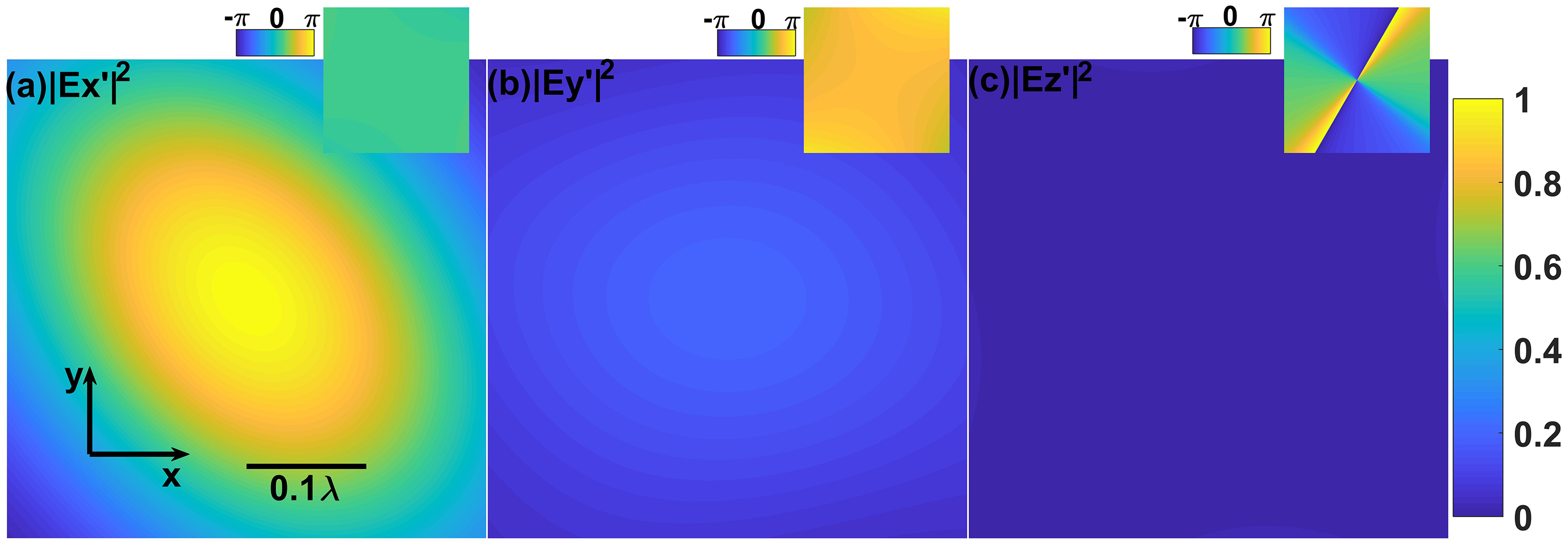}
	\caption{Intensity and phase (insets) profiles on the
focal plane when $f=-ig_{-1}=g_1/\sqrt 3$.}
	\label{Fig:EP}
\end{figure}

  Figure \ref{Fig:EP} is an example of  elliptical polarization with  $f=-ig_{-1}=g_1/\sqrt 3$. In order to display the polarization clearly, we use an orthogonal transformation:
  
  \begin{equation}
      \begin{aligned}
      \left[\begin{array}{ccc}E_{x'} \\  E_{y'} \\E_{z'} \end{array}\right]= \left[\begin{array}{ccc} \frac{\sqrt{3}}{\sqrt{10}}\ \frac{-2}{\sqrt{10}} \frac{\sqrt{3}}{\sqrt{10}} \\  \frac{1}{\sqrt{2}} \ \frac{0}{\sqrt{2}} \ \frac{-1}{\sqrt{2}} \\\frac{1}{\sqrt{5}}\ \frac{\sqrt{3}}{\sqrt{5}}\ \frac{1}{\sqrt{5}} \end{array}\right]\left[\begin{array}{ccc}E_x \\  E_y \\E_z \end{array}\right].
      \end{aligned}
  \end{equation}
  As shown in Fig. \ref{Fig:EP}, $|E_{z'}|^2$ is nearly zero. The  phase retardation between  $x'$-component  and  $y'$-component are about $\pi/2$, but their amplitudes are different. So the focal field is elliptically polarized.  The minimum, median and maximum TFWHM are 0.34, 0.37 and 0.40$\lambda$, respectively. The polarized homogeneity is 0.988 (R=0.22$\lambda$).

\section{Conclusion}
In conclusion, we presented an approach to generate an arbitrary 3D polarization by tightly focusing vector beams. The size of the focal spots is 0.32 $\sim$ 0.44$\lambda$. The polarized homogeneity  is always large than 0.97.

\bibliography{new}

\end{document}